# Low-cost spectrogram based counterfeit medicine and contaminated food detection

By A. K. Mishra and M. H. Essop
University of Cape Town (akmishra@ieee.org )

*Abstract*—Contaminated substances such as counterfeit medication and food contaminated with pesticide residue is a pandemic of utmost urgency. Spectroscopy and chromatography methods are often used to test for contaminants. However, they are expensive and complex. There is a need for an inexpensive and easy-to-use technological solution to detect counterfeit medicines or contaminated food. Following the philosophy of application specific instruments, we present a hacked visible spectrometer based contaminated substance detector using machine learning. Our results show that a low-cost method of identifying contaminated substances is achievable with very high accuracy.

## I. INTRODUCTION

Counterfeit medicines have become one of the world's fastest growing industries, becoming a global pandemic as it jeopardizes patient health and recovery. It has been reported that the use of counterfeit drugs has been linked to an increase in morbidity, drug resistance, and in the most severe cases, death [1]. A similar problem is that of pesticide residue ingested from fruit and vegetables which has been linked to an increase in cancer, as well as negative effects on the reproductive, immune and nervous systems [2].

Various methods to prevent the trade of counterfeit medicines have been put in place by government; such as regulating the manufacture, export, import, supply and distribution as well as the sale of the medicines itself [3] . However there is little to no double checking of the medications quality by the tertiary suppliers themselves, such as the pharmacists or doctors in the field, whom are ultimately the agents of distribution of medicines to the patients [4].

Self-absorption based technologies have been foreseen to be crucial in future engineering solutions [5]. Advances in counterfeit medicine detection and pesticide detection have been made using techniques such as spectrometry, radio-frequency identification and chromatography [6], [7]. The use of spectrometry as a technique for counterfeit medicine detection has shown promising and reliable results. However these methods require equipment beyond the financial means of developing countries. In addition, these methods require prerequisite training by medical workers to operate the state of the art technology in the field. Another major drawback of these approaches is the lack of accuracy when investigating murky or colloidal solutions thus requiring further sample preparation and filtering.

Based on the high cost and training required to operate commercial and laboratory methods of detecting contaminated substances, the development of a simple low-cost counterfeit medicine and contaminated fruit detector is essential for developing countries. In our endeavour to find a solution for this challenging issue we hypothesized that the detection of the quality of a substance does not have to be through the identification of constituent compounds. This way of instrument development [8] assumes that we can directly detect the event of interest using inexpensive sensors and using machine learning. Previous work by the first author has shown promising results using lowcost spectroscopy techniques [9]. The proposed device operates on the bases that the visible spectrum produced by light that passes through a sample represents the chemical fingerprint of that sample and can be used to identify its ingredients.

In the current work we present the design and implementation of a visible spectroscopy based counterfeit medicine detector using machine learning. This technique allows reliable predictions to be made without requiring the necessary knowledge of pharmaceutics. The device was proposed as a solution to combat medication adulteration. The system was also used for the detection of lethal doses of pesticide residue in fruit juice which poses a threat in nearly 70% of all commercial fruit sold [10]. The spectrometer diffracts the continuous spectrum produced by an incandescent lightbulb throughout the visible region of the electromagnetic spectrum, $390-700nm$. The incident light travels through the entrance slit where it is then scattered by the sample of interest. Finally the scattered light is diffracted by the DVD diffraction grating in order to produce a continuous visible spectrum which is then captured by the CMOS detector, a logitech C170 webcam. The acquired spectrum images have then been cropped to the dimensions $150\times150\times3$ pixels and preprocessed, as required by the ML algorithms. The spectrometer costs a total of R600 ($\approx$\$45) and achieves a spectral resolution of 2.1nm/pixel. We demonstrate the development of a hacked visible spectroscopy based contaminated substance detector using machine learning, achieving remarkable accuracies of up to 100% thus proving that the concept is not only possible but feasible.

Rest of the paper is organized as follows. Section II describes the system requirements and explores existing approaches. Section III discusses the design of the experiments and system including hardware, software, ML and experiment. Section IV discusses the results acquired per experiments. Finally, Section V concludes with final remarks and discussion of further work.

## II. AIM AND REQUIREMENT ANALYSIS

### A. Problem Specification

According to the World Health Organisation (WHO), "A counterfeit medicine is one which is deliberately and fraudulently mislabelled with respect to identity and/or source. As such, counterfeiting can apply to both branded and generic products" [4]. As reported by WHO there are three major types of counterfeit medicine:

1) Placebo - Drugs with no Active Pharmaceutical Ingredients (API).
2) Generic - Drugs where one or more of the costly API's have been replaced by cheaper alternatives.
3) Dilute - Drugs that contains lower concentrations of the API.

Pesticide residue in fruit and vegatables has also been of growing concern as producers aim to increase the shelflife og their produce. Pesticide residue is a carcinogen and the ingestion of contaminated foods has negative consequences on the reproductive, immune and nervous system [10]. The Environmental Working Group (EWG) has reported that 70% of fruits and vegtables grown for commercial use contain 230 different pesticeds and residue thereof [11]. Apple, spinach, grapes, strawberry and cherries have topped the list of contaminated foods.

### B. Existing approaches

A method has been proposed that centres around the development of a detection device that determines if an object is a counterfeit pharmaceutical product as well as being able to identify counterfeit packaging [6]. The device consists of multiple LED's, infrared, visible and ultra-violet light sources, which emits light at different wavelengths. The device houses an image acquisition device, a CCD array, which receives the returning light. This approach involves selecting a wavelength range of light, exposing the counterfeit drug to the light and then comparing it with the appearance of an authentic drug to the same light. If the appearance differs then the product being analyzed is declared a counterfeit, where the product is either a tablet, capsule or package. Raman Spectroscopy, using visible light, has also shown promising results for the identification of pesticide residue in fruit [7].

Another study has applied a convolutional neural network (CNN) on vibrational spectroscopy data [12]. The study implements a CNN architecture with a single convolutional layer is implemented in order to extract features free of preprocessing methods. A novel regularization term is implemented which smooths the input data enabling the CNN model to easily adapt to different unseen spectral inputs. It has been demonstrated that a CNN with one hidden convolutional layer performs significantly better than a Partial Least Squares Regression model with both raw and preprocessed data, where cross validation was used to find the optimal preprocessing method.

As such these methods are far too expensive and complex to be implemented in rural and developing areas with the biggest limiting factor being the pre-requisite training required by medical workers in the field.

Work done by the first author on a novel system for the identification of counterfeit drugs has served as inspiration for the design of the spectrometer implemeted in our device [9].

### C. Requirement Analysis

Thus the aim is to develop a low-cost spectrometer that allows light to be passed through a sample, diffracted into its individual wavelengths and then recorded. Utmost care is taken to ensure that the instrumentation is of good enough standard as to achieve acceptable spectral resolution while remaining affordable.

- Low-cost: The system is to be designed with lowcost components that are easily available and replaceable.
- Modularity: The system components are to be modular and independent. This allows for easy repair and assembly of the working system.
- Performance: The system should be accurate and robust in its recognition of counterfeit medication and contaminated fruit. The device should be capable of identifying counterfeit medication and contaminated fruit juice with 90% accuracy using spectroscopy methods.
- User-friendliness: The system should be convenient to use with little-to-no experience in the field.
- Robustness: The device should be capable of working predictably in different lighting and temperature environments.

This approach aims to deliver a device to be operated with little-to-no prior training and is capable of producing accurate and reliable results, thus being accessible to developing countries where the medical workforce is scarce.

### III. SYSTEM DESIGN

#### A. System Overview

After much research and design, Spector was created. Spector is a hacked spectrometer that uses a machine learning algorithm to classify a substance as authentic or contaminated. The device consists of four major independent modules, thus allowing them to be developed in parallel before being integrated into a working system. The four modules are: the Visible Spectrometer, the experiment and data acquisition module, the data pre-processing module and the machine learning (ML) module.

The four modules are designed to be implemented in series with one another, as indicated in figure 1. Thus the spectrometer produces the visible spectrum of the sample prepared in the experiment module, which is the 390 700nm region of the EM spectrum. The sample is prepared in distilled water which is then placed in the spectrometer using the cuvette. The incident light passes through the sample to be identified which ultimately affects the energy and colour of the light due to absorption and scattering processes. The scattered light then passes through an inlet slit where it meets the diffraction grating and image sensor which acquires the spectrum data.

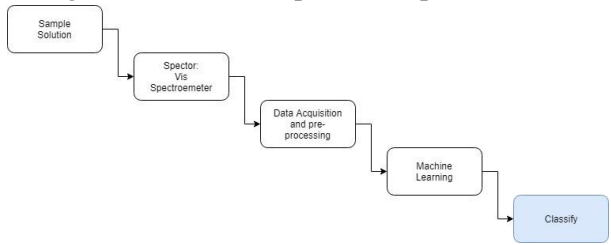

Fig. 1. System operation

Each sample scatters and absorbs the incident light uniquely essentially providing a chemical fingerprint of that sample, thus the acquired data is used to train the machine learning algorithms to classify whether a substance is contaminated or not based on the visible spectrum it produces.

#### B. Experiment Design

The medication used and its ingredients are summarised below:
- 600mg Panado tablet - 500mg Paracetamol
- 700mg Ibumol tablet - 200mg Ibuprofen, 350mg paracetamol
- 600mg Sinuend tablet - 200mg paracetamol, 20mg caffeine, 6mg ephedrine hydrochloride, 2mg Chlorpheniramine maleate
- 500mg Azithral Antibiotic - 500mg Azithromycin dihydrate
- 700mg Cefadroxil Antibiotic - Cafadroxil gemihydrate, cefadroxil monohydrate

The authentic, placebo and generic medication samples were prepared using 400mg of the crushed tablet dissolved in 40ml of distilled water. Where the diluted sample contained 75%, 50%, 25% and 0% of the crushed substance while having its mass preserved with white sugar. The models are trained to classify Azithral and Cefadroxil antibiotics, paracetamol, Ibuprofen and any diluted or placebo forms thereof. All prepared samples are indistinguishable to the human eye.

The pesticide experiment is conducted using four samples. The control sample contained 100% pure juice. The three experiment samples contained different levels of doom pesticide, also all indistinguishable by the human eye.

Finally the cuvette is placed in the spectrometer and the data is captured by the computer. 1000 images are taken per sample in different locations as to ensure that changes in ambient light and temperature are accounted for.

#### C. Hardware Design

The spectrometer is the instrument used to acquire the visible spectrum of the sample of interest. When designing the spectrometer shell care was taken as to prevent any ambient light from leaking through which could prove detrimental to the quality of the captured spectrum. Thus a 100W incandescent light bulb, connected to a 220V external supply, is used as the light source due to the continuous spectrum it produces in the visible region. The lightbulb also has a high intensity light thus rendering the effects of leaked ambient light obsolete. Special care was taken when selecting the cuvette, diffraction grating and optical sensor components as to ensure that all components allowed the transmission and detection of visible light. Thus a 3mm glass cuvette is used as it allows the transmission of the visible spectrum which is then diffracted by a 1350line/mm DVD diffraction grating and captured by the Logitech C170 Webcam [13], [14].

The spectrometer is successful in producing a focused and clear diffraction spectrum of the incident light spanning the entire visible region of the EM spectrum. The images are cropped to the dimensions 150x150 as to only contain the visible spectrum, as shown in figure 3.

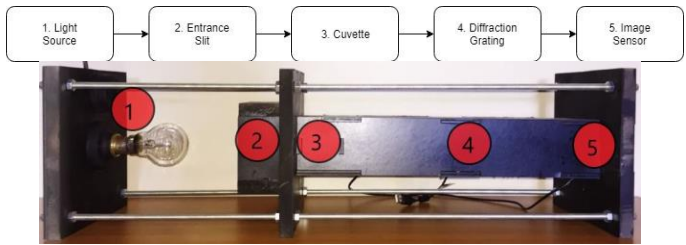

Fig. 2. Visible spectrometer device operation

## D. Machine Learning

The ML models are trained and implemented in Python 3.0 using the Tensorflow, Keras and Sklearn packages due to their high level abstraction and various ML algorithms and methods. All captured data and trained models are stored locally with the largest model occupying near 80MB of memory whereas each image occupies 5kB.

Before being passed into the machine learning algorithm, the raw pixel values are pre-processed to be used as features for training the machine learning algorithm. In order for the acquired data to be implemented for supervised machine learning they were labeled according to their API name. The raw pixel data, ranging from 0 to 255, is then extracted and stored as a one dimensional array of pixel values to be used by the ML algorithm for training or prediction. Finally the data is in an acceptable format and representation and the ML models can now be trained.

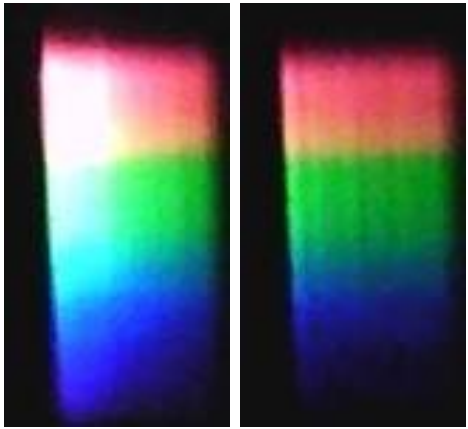

Fig. 3. Visible spectrum of the Authentic (left) and Generic (right) samples as acquired via the spectrometer prior to pre-processing

A convolutional neural network (CNN), a support vector machine (SVM) and a logistic regression model are implemented for the purpose of binary classification of authentic or contaminated substance. The preprocessed spectrum arrays are then used as the input data for training and validating the machine learning algorithms. A linear regression model is implemented for the purpose of identifying the percentage by which a substance is diluted, where the model is trained on the images raw pixel and integer label data.

The SVM and regression models use the raw pixel data to fit the model thus defining a mathematical function that best separates the data. The difference in their performance is due to their operation. The SVM defines a hyperplane that best separates the data using a linear function with a maximum hard margin. Whereas the regression models finds the mean squared error (MSE) line of best fit that best describe the data.

The CNN model is more abstract as it uses three convolution and pooling layers which extract higher and higher features of the image which is then used for training and predictions. That being said, the CNN model is computationally expensive due to the sheer number of trainable parameters [15].

## IV. RESULTS AND DISCUSSION

All trained models achieved greater than 90% validation accuracy indicating that the models are well generalized to make predictions on new unseen test data. The models are trained using 1000 images of each sample and are validated through the process of cross validation, with a 80:20 train and test data split.

### A. Placebo and Generic Experiments

The logistic regression, SVM and CNN models is implemented for the purpose of classifying an authentic medicine from a placebo/generic. The results for these models are consistent across both experiments and as such will be discussed as one.

The SVM and logistic regression models are the best performing classifiers both achieving 99 to 100% accuracy, as shown in figure 4. Their near identical results are attributed to the data being linearly separable due to the large number of pixel features.

The CNN model is accurate in identifying placebo medication as shown in figure 5 achieving a validation accuracy accuracy of 99% on the 15th epoch.

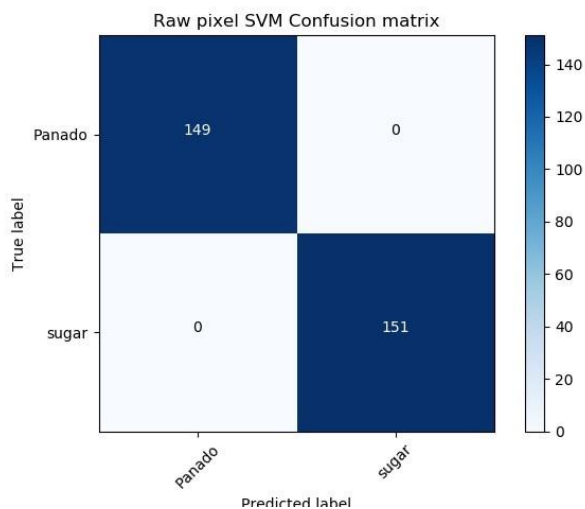

Fig. 4. SVM model confusion matrix on the predictive performance for identifying placebo medication

*B. Dilute Experiment*

The linear regression model achieved an accuracy of greater than 97% with a correlation coefficient of 0.99, thus the model is a near perfect for the purpose of predicting by what percentage a substance has been diluted. Table I represents the performance for all medication sampled.

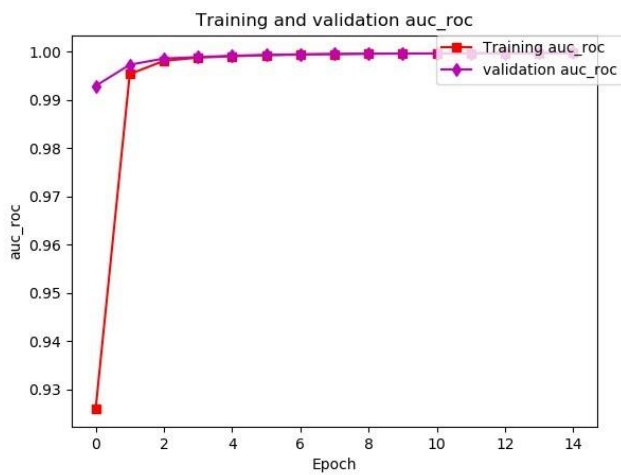

Fig. 5. CNN model training and validation ROC AUC plot for 15 epochs, for the purpose of identifying placebo and generic medication

TABLE I
LINEAR REGRESSION MODEL VALIDATION ROC _AUC PLOT FOR 15 EPOCHS, FOR THE PURPOSE OF IDENTIFYING PLACEBO AND GENERIC MEDICATION

| Medication | Accuracy % | Correlation Coefficient |
|---|---|---|
| Panado | 98.61 | 0.99 |
| Sinuend | 99.96 | 1 |
| Azithral | 97.1 | 0.97 |
| Cefadroxil | 99.53 | 1 |

The Logistic regression model and the SVM model achieved near identical results, for reasons previously mentioned. As shown in figure 6 the models are used for multi-class classification labelling the placebo below a specific level of diluteness.

The Dilute Panado experiment also sees the greatest loss. This could be due to the fact that Panado powder dissolved in water displays photo-luminescent effects when the concentration is above 75% thus making it difficult to distinguish between pure paracetamol and a sample that has been diluted by 25%.

*C. Pesticide Residue Experiment*

The SVM-SVC and Logistic Regression model used for the identification of counterfeit medication has been used for the detection of pesticide in juice. As such the only change was in the datasets used to fit the respective models.

Table II indicates the accuracy, Hinge loss and MCC score for the respective models per experiment. The MCC score of both models indicate that the models are well fit to generalize on new unseen data thus making great classifiers.

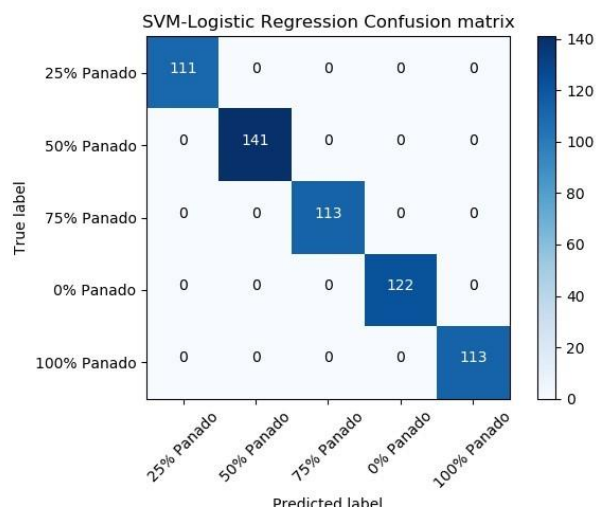

Fig. 6. Logistic Regression model confusion matrix on the predictive performance for identifying dilute medication

TABLE II
SVM AND LOGISTIC REGRESSION MODELS PERFORMANCE
METRICS ON THE PREDICTIVE PERFORMANCE FOR IDENTIFYING FRUIT
JUICE CONTAINING PESTICIDE RESIDUE

| Medication | Performance Metrics | | | | | |
|---|---|---|---|---|---|---|
| | SVM | | | Logistic Regressor | | |
| | ACC % | Loss | MCC | ACC % | Loss | MCC |
| Apple | 100 | 0 | 1 | 100 | 0 | 1 |
| Cranberyy | 99.58 | 0.42 | 0.99 | 97.71 | 2.29 | 0.97 |

The CNN model achieves a validation accuracy score of above 95%, thus indicating that the system generalizes well to new unseen data.

It is noted that the apple juice experiment performed best. This is thought to be due to the fact that apple juice is more transparent than cranberry juice and as such allows more light through, thus increasing the quality of the data recorded.

*D. Discussion of Results*

The spectrometer captured the entire visible region of the electromagnetic spectrum, $390 - 700nm$, and the spectrum was used as data for training and fitting the machine learning models. Thus the project demonstrates that a spectrometer developed with a budget of R600 is capable of achieving a spectral resolution of $\approx 2.1 nm/pixel$.

A Convolutional Neural Network, support vector machine and a Logistic regressor has been implemented for the purpose of classifying a counterfeit medicine from an original as well as for the purpose of identifying pesticide residue in juice. The CNN model achieved the lowest accuracy for the purpose of identifying a contaminated substance with a score of 91% whereas an accuracy of $100 - 99\%$ was achieved by the logistic regression and SVM models.

The results achieved indicate that a machine learning model can be trained on a relatively small dataset of 2400 - 3000 images and achieve a near perfect accuracy in order to determine the integrity of medication and juice using visible spectrum data. All machine learning models implemented generalised well to new unseen as high validation accuracy scores were achieved, closely matching the training scores.

The Linear regression model has been fit for the purpose of identifying the percentage of the active pharmaceutical ingredient is present within a sample. The lowest and highest accuracy achieved on an unseen test set is 97.1% and 99.96% respectively. The lowest and highest correlation coefficient achieved is 0.97 and 1, as such indicating that a near perfect predictor has been achieved.

The results achieved by the linear regression model indicates that machine learning techniques can be fit using a relatively small dataset of 3000 images to detect the effect of medication adulteration.

The support vector machine yielded the best results for the purpose of identifying a contaminated substance such as a generic, placebo or diluted medicine as well as pesticide residue in juice as previously shown. During experimentation it was noted that the Logistic regression model had the shortest training and prediction time amongst all the models. With regards to memory consumption of the trained/fit classification models the SVM model is largest at 190MB, the CNN at 9.7MB and the logistic regression model at 2.7MB. The linear regression model occupies 0.6MB. As such there is a trade off between the predictive accuracy of a classification model and the space it occupies in memory. The CNN took the longest to train as it has the most parameters, as such an improved accuracy can possibly be achieved by training the model on a larger dataset over more epochs.

V. CONCLUSION

In this work we demonstrated the use of an inexpensive spectroscopy based system to detect counterfeit medicines and adulterated food. Some medication samples such as the the Panado, Cefadroxil and pesticide showed photo-luminescent effects thus producing a narrow diffraction spectrum of high intensity. However the Azithral and Ibumol dispersed the incident light more than the other medication thus producing a wider spectrum of lower intensity. Although undesired, these effects were unique to each medication and have contributed to the ability to distinguish them from one another.

The material cost of the system was around $50. Within experimental error the system is also robust as consistent results have been achieved across all machine learning models and experiments.

*A. Potential Improvements and Further Work*

It is a trivial hypothesis that the addition of scattering data over a larger bandwidth would bolster the performances. This requires sensors which can capture spectrum in ultra violet, visible and infrared bands. A larger spectrum will lead to an improvement in the achieved spectral resolution thus providing more data points, features, for fitting the machine learning models. An increase in spectral resolution will allow the measurement of light intensity at specific wavelengths thus allowing the chemical fingerprint of a substance to be determined using ML techniques, subsequently allowing the ML models to be trained for more than substance classification.

In order to acquire a larger spectrum an Infrared quartz cuvette is to be used as it has a bandpass that spans the UV - NIR region of $220 - 3500 nm$. A microscopic slide can be used in place of the cuvette as to only hold a single drop of the sample solution. The thin sample layer between the surface of glass will prevent the presence of colloids to produce a spectrally pure image of the sample, allowing the system to be implemented for microscopic analysis such as detecting pathogens or disease in blood.

A motorized stage or light source will enable the capture of multiple points per sample thus allowing us to further account for any variations in the sample solutions consistency. A motorized stage will increase the size and quality of the dataset implemented for training the machine learning models.

By acquiring a larger dataset that includes multiple medication as well as its respective generics and placebos, a single larger multi-classification SVM model can be developed for the identification and classification of multiple drugs of interest. A larger dataset where the APIs and all its excipients have been labelled will enable a machine learning model to be developed where the exact composition of a medication or juice can be identified.

Implementation of the above improvements along with the addition of features for seamless user-interface will make this solution of much wider adoption and applicability.

## ABOUT THE AUTHORS

**Amit Kumar Mishra** (akmishra@ieee.org) is an Associate Professor with the Department of Electrical Engineering at the University of Cape Town.

**Mohamed Hoosain Essop** (mh.essop@gmail.com) is a Tech and Business Integration Analyst with Accenture South Africa.

## REFERENCES

[1] R. Cockburn, P. N. Newton, E. K. Agyarko, D. Akunyili, and N. J. White, "The global threat of counterfeit drugs: why industry and governments must communicate the dangers," *PLoS medicine*, vol. 2, no. 4, p. e100, 2005.

[2] B. Lozowicka, "Health risk for children and adults consuming apples with pesticide residue," *Science of the Total Environment*, vol. 502, pp. 184–198, 2015.

[3] D. Shore, "Counterfeit medicines and the need for a global approach," *Eur Pharm Rev*, vol. 20, pp. 10–13, 2015.

[4] W. E. C. on Specifications for Pharmaceutical Preparations and W. H. Organization, *WHO Expert Committee on Specifications for Pharmaceutical Preparations: Thirty-ninth Report*, vol. 39. World Health Organization, 2005.

[5] J. Carvalko, "Self absorption: Where will technology lead us?," *IEEE Consumer Electronics Magazine*, vol. 5, no. 1, pp. 120–122, 2016.

[6] N. Ranieri, M. R. Witkowski, W. G. Fateley, and R. Hammaker, "Device and method for detection of counterfeit pharmaceuticals and/or drug packaging," Oct. 25 2016. US Patent 9,476,839.

[7] S. Dhakal, Y. Li, Y. Peng, K. Chao, J. Qin, and L. Guo, "Prototype instrument development for non-destructive detection of pesticide residue in apple surface using raman technology," *Journal of Food Engineering*, vol. 123, pp. 94–103, 2014.

[8] A. K. Mishra, "Application specific instrumentation (asin): A bio-inspired paradigm to instrumentation using recognition before detection," *arXiv preprint arXiv:1611.00228*, 2016.

[9] A. K. Mishra and R. Janowski, "Speky: A diffraction based application specific instrumentation (asin) counterfeit medicine detection system," in *Computer, Communication and Signal Processing (ICCCSP), 2017 International Conference on*, pp. 1–5, IEEE, 2017.

[10] P. Cozma, L. C. Apostol, R. M. Hlihor, I. M. Simion, and M. Gavrilescu, "Overview of human health hazards posed by pesticides in plant products," in *E-Health and Bioengineering Conference (EHB), 2017*, pp. 293–296, IEEE, 2017.

[11] Y. Hu, Y.-H. Chiu, R. Hauser, J. Chavarro, and Q. Sun, "Overall and class-specific scores of pesticide residues from fruits and vegetables as a tool to rank intake of pesticide residues in united states: a validation study," *Environment international*, vol. 92, pp. 294–300, 2016.

[12] J. Acquarelli, T. van Laarhoven, J. Gerretzen, T. N. Tran, L. M. Buydens, and E. Marchiori, "Convolutional neural networks for vibrational spectroscopic data analysis," *Analytica chimica acta*, vol. 954, pp. 22–31, 2017.

[13] "The cd rom spectroscope." http://astro.u-strasbg.fr/koppen/spectro/spectroe.html. Accessed: 2018-09-26.

[14] "Transparent materials comparison." https://rayotek.com/techspecs/material-comparisons.htm. Accessed: 2018-09-30.

[15] S. AL-AZAWI, "Understanding of a convolutional neural network," in *2017 International Conference on Engineering and Technology (ICET)*, IEEE, 2018.